\documentclass{article}

\usepackage{PRIMEarxiv}

\usepackage[utf8]{inputenc} 
\usepackage[T1]{fontenc}    
\usepackage{hyperref}       
\usepackage{url}            
\usepackage{booktabs}       
\usepackage{amsfonts,amsmath,amssymb}       
\usepackage{nicefrac}       
\usepackage{microtype}      
\usepackage{lipsum}
\usepackage{fancyhdr}       
\usepackage{graphicx}       
\usepackage{graphics}
\graphicspath{{media/}}     

\usepackage{cite}
\usepackage{algorithmic}
\usepackage{textcomp}
\usepackage{bm}
\usepackage{upgreek}
\usepackage[ruled,vlined,linesnumbered]{algorithm2e}
\usepackage{caption}
\usepackage{subcaption}

\def\pm{PM$_{2.5}$ }

\title{Sensor data-driven analysis for identification of causal relationships between exposure to air pollution and respiratory rate in asthmatics}

\author{
  D K Arvind \\
  Centre for Speckled Computing \\
  University of Edinburgh \\
  Edinburgh\\
  \texttt{dka@ed.ac.uk} \\
   \And
  S Maiya \\
  Centre for Speckled Computing \\
  University of Edinburgh \\
  Edinburgh\\
  \texttt{smaiya@ed.ac.uk} \\
}

\begin{document}
\maketitle

\begin{abstract}
According to the Lancet report on the global burden of disease published in October 2020, air pollution is among the five highest risk factors for global health, reducing life expectancy on average by 20 months. This paper describes a data-driven method for establishing causal relationships within and between two multivariate time series data streams derived from wearable sensors: personal exposure to airborne particulate matter of aerodynamic sizes less than 2.5$\bm{\upmu}$m (\pm) gathered from the Airspeck monitor worn on the person and continuous respiratory rate (breaths per minute) measured by the Respeck monitor worn as a plaster on the chest. Results are presented for a cohort of 113 asthmatic adolescents using the PCMCI+ algorithm to learn the short-term causal relationships between lags of \pm exposure and respiratory rate. We consider causal effects up to a maximum delay of 8 hours, using data at both a 1 minute and 15 minute resolution in different experiments. For the first time a personalised exposure-response relationship between \pm exposure and respiratory rate has been demonstrated to exist for short-term effects in asthmatic adolescents during their everyday lives. Our results lead to recommendations for work on specific open problems in causal discovery, to increase the feasibility of this approach for similar epidemiology studies in the future.
\end{abstract}


\section{Introduction}
Air pollution in the form of airborne particulate matter, nitrogen dioxide and ozone is the fifth largest contributor to the global burden of disease and was estimated to account for more than five million deaths in 2017 \cite{b_1}. 
Vulnerable groups may be at an even greater risk of severe health decline due to air pollution exposure \cite{b_2}. These groups include the elderly, children and those with chronic underlying conditions such as coronary heart disease, asthma and COPD (chronic obstructive pulmonary disease) - an umbrella term for lung conditions such as emphysema, chronic bronchitis and upper airways obstruction. In addition, other groups may be exposed to higher levels of indoor (e.g. solid cooking fuel users) or outdoor (e.g. those living near busy traffic routes or those in specific occupational or socioeconomic groups \cite{b_3}) air pollution than most, and the impact on their health is of concern.

Health risks due to air pollution are often presented in terms of attributable deaths, cases of disease, years of life lost, disability-adjusted life years or change in life expectancy attributable to absolute exposure or change in exposure to air pollution. This analysis utilises concentration-response functions (CRFs) also known as dose-response functions (DRFs) or exposure-response relationships (ERRs) which quantify the units of health impact per concentration unit of a particular air pollutant. For airborne particulate matter (which can be sourced from combustion processes, detritus from wear and tear of brakes and tyres, sea salt, re-suspension of soil) this is commonly measured from the mass-concentration ($\upmu$g/m$^3$) of particles smaller (in diameter) than 1$\upmu$m (PM$_{1}$), 2.5$\upmu$m (PM$_{2.5}$) or 10$\upmu$m (PM$_{10}$).

\begin{figure}[h]
    \centering
    \begin{subfigure}{0.24\textwidth}%
        \includegraphics[width=\textwidth, keepaspectratio=True]{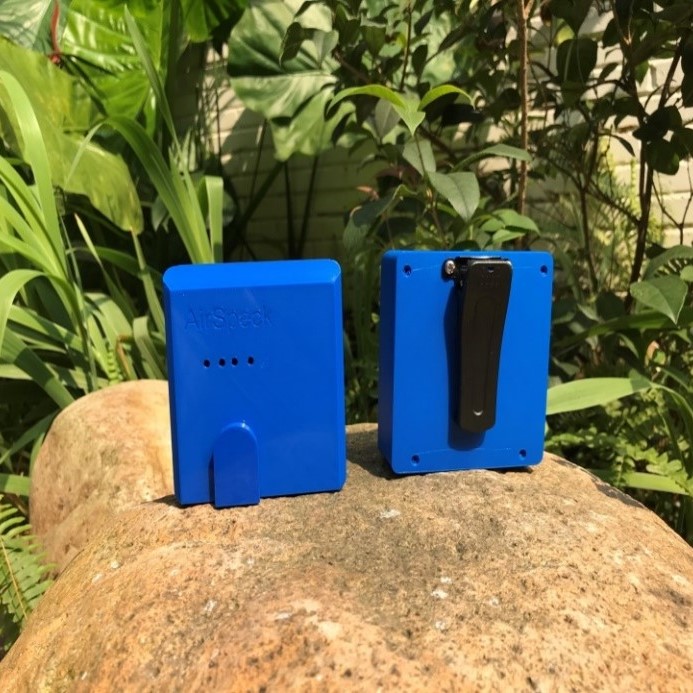}%
    \end{subfigure}%
    \hspace{10em}
    \begin{subfigure}{0.24\textwidth}%
        \includegraphics[width=\textwidth, keepaspectratio=True]{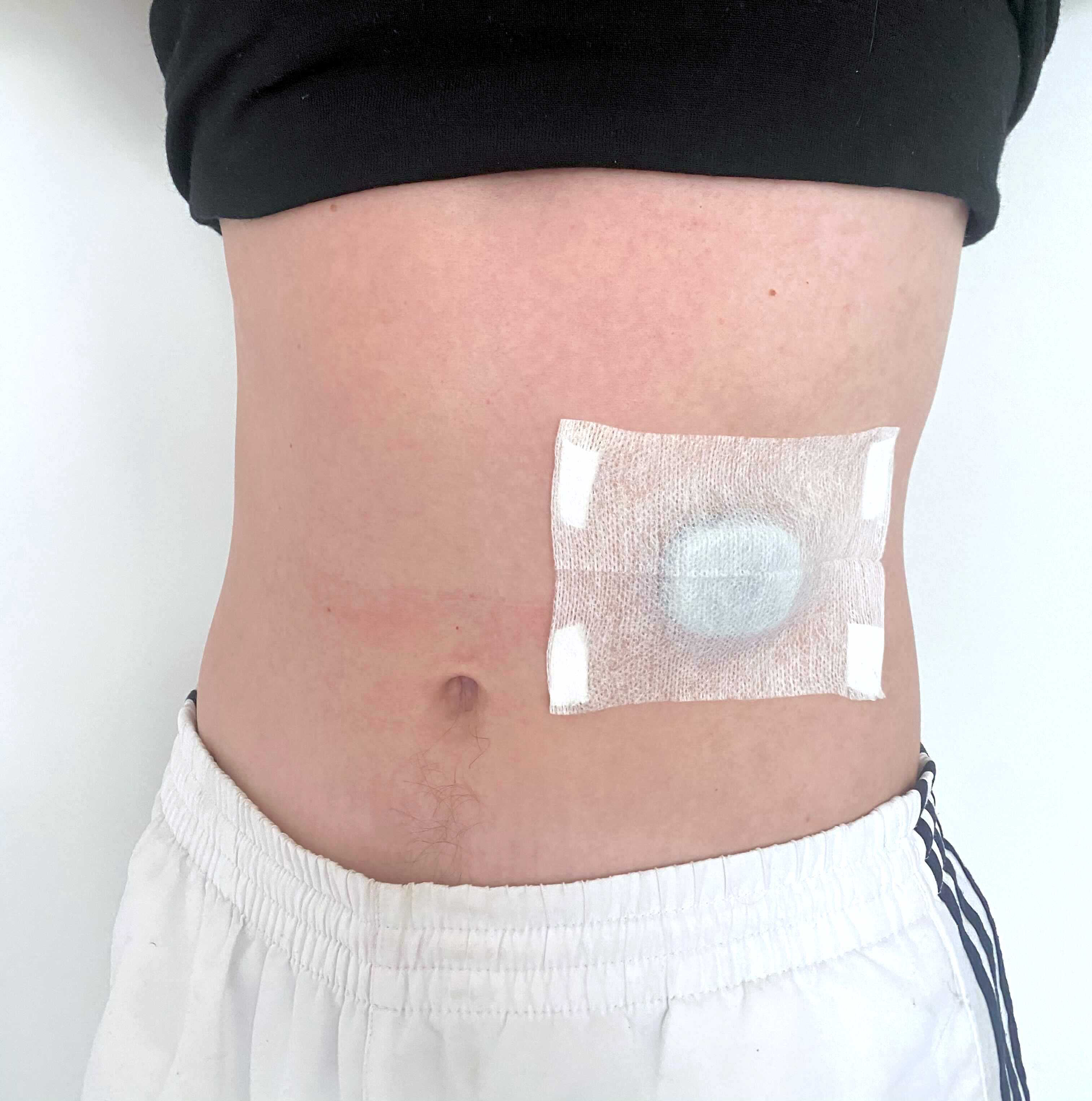}%
    \end{subfigure}%
    \caption{(L) The Airspeck personal device, used for continuous monitoring of air pollution exposure, temperature and relative humidity. (R) The Respeck device, for continuous monitoring of breathing rate and physical activity.}
    \label{fig:airspeck/respeck}
\end{figure}

\begin{figure}[h]
    \centering
    \includegraphics[width=0.4\textwidth]{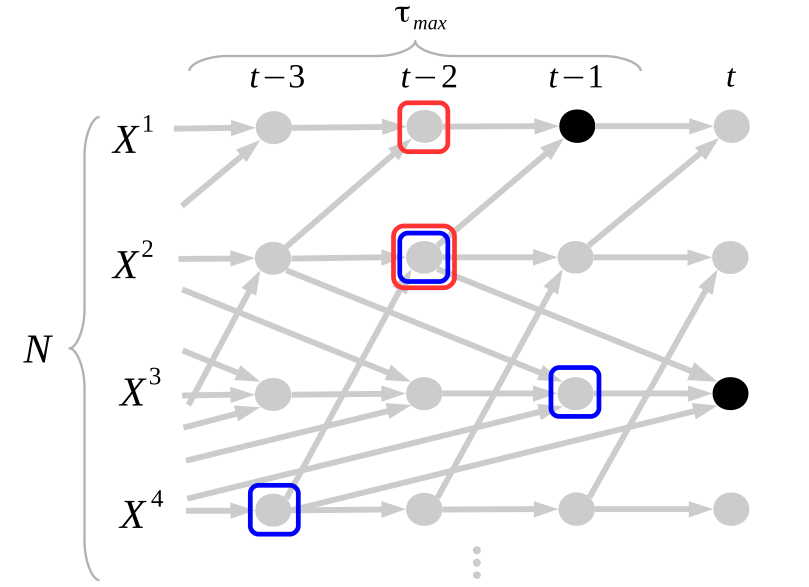}
    \caption{Figure from \cite{b_17}, showing a section of a full causal graph. This illustrates direct causal relationships between lags of variables in the system under study.}
    \label{fig:causal_graph}
\end{figure}

In contrast, we seek to identify the \textit{short-term} (on the order of minutes to several hours) effects of exposure to air pollution on a personal level based on continuous monitoring of subjects using wearable sensors \cite{b_10}. These sensors send data wirelessly to a mobile phone, where it is time-stamped, and transmitted to a cloud-based server for storage, curation and analysis. Figure \ref{fig:airspeck/respeck}-L shows the Airspeck sensor \cite{b_12} which when clipped to the person's clothing records personal \pm exposure, temperature and relative humidity. Figure \ref{fig:airspeck/respeck}-R shows the Respeck sensor \cite{b_11, b_21}, worn as a plaster on the chest to record the subject's respiratory rate and respiratory flow/effort, and physical activity. A synchronised multivariate time series is constructed using the data from both sensors recorded at minute-level intervals. The PCMCI+ algorithm \cite{b_14, pcmciplus} is applied to learn the causal structure of the data-generating system and produce a causal graph (or causal network) - a directed acyclic graph (DAG) with nodes for all variables at all time lags and edges drawing direct cause-effect relationships between pairs of lags of variables (see Figure \ref{fig:causal_graph}). The analysis presented in this paper is based on a dataset collected contemporaneously from Airspeck and Respeck sensors worn by asthmatics for a period of 48 hours each, resulting in a personalised causal graph for each subject.

PCMCI+ is a constraint-based causal discovery algorithm \cite{b_15} which learns the causal structure of a multivariate time series, testing for causal relationships between pairs of lags of variables and producing a causal graph such as the one shown in Figure \ref{fig:causal_graph}. PCMCI+ and similar methods require an explicit selection of a maximum lag-length parameter, $\tau_{max}$, before running: this is defined as the largest possible delay between a cause and effect. For Figure \ref{fig:causal_graph} the value $\tau_{max}$ set to 3 could be three minutes, hours, months, etc. depending on the resolution of the time series $\mathbf{X} = \{X^1, X^2, X^3, X^4\}$. Each grey arrow can be read as a direct causal relationship; for example $X^1_{t-2}$ and $X^2_{t-2}$ directly cause $X^1_{t-1}$ (red); $X^2_{t-2}$, $X^3_{t-1}$ and $X^4_{t-3}$ directly cause $X^3_t$ (blue).

Given a variable $X^i_t$ in a causal graph, the direct causes of that variable are referred to as causal parents, with the set of all causal parents denoted $\mathcal{P}(X^i_t)$. The effect of changes in a variable $X^p$ on another variable $X^q$ can be investigated by looking at the subset of $\mathcal{P}(X^q_t)$ of lags of $X^p$, which can be used to model an exposure-response relationship.

It should be noted that estimating the causal graph itself is expensive - the search space of possible DAGs is exponential in the number of variables. Most constraint-based causal discovery methods follow one of two paths: either initialise an empty graph and recursively add edges or initialise a fully-connected undirected graph and recursively remove edges. The latter approach is taken in the PC algorithm \cite{b_16}, a well-known causal discovery algorithm which runs in exponential time in the worst case, but polynomial time if the DAG is sparse, which is a reasonable assumption to make in a majority of real cases. Edges of the (initially) fully-connected graph are removed using a test for conditional independence\footnote{Conditional independence is closely related to the information-theoretic quantity of conditional mutual information, from which a particular mathematical definition of causality is obtained \cite{b_18}.}, with the choice of test dictating whether the method has strength in determining linear or nonlinear causal relationships. A modified PC algorithm is used in PCMCI+ which follows the same general approach. This modification addresses an issue encountered when applying the PC algorithm to time series: the original algorithm was found to incorrectly control for some highly autocorrelated links (false positives). The modification in PCMCI+ results in a better false positive discovery rate in both linear and nonlinear settings.

The novel contributions of this paper are in developing and critically evaluating a method for establishing causal relationships between personal \pm exposure and the respiratory rate of asthmatics during their everyday lives, and in demonstrating individual exposure-response relationships for this effect in the short-term (up to eight hours).

\subsection{Related Work}
Schwartz et al. \cite{b_4, b_5} investigated through retrospective studies the association of concentrations of coarse inhalable particles (PM$_{10}$) and ozone levels with Detroit hospital admissions for persons aged over 65 years during the period 1986 to 1989. Using a Poisson regression model controlling for seasonal trends, ambient and dew point temperatures, associations were established between both PM$_{10}$ and ozone with hospital admissions for pneumonia and COPD exacerbations, hinting at a possible causal relationship for both pollutants. Anenberg et al. \cite{b_6} estimated that 9-23 million and 5-10 million annual asthma emergency room visits (ERVs) during 2015 across 54 countries and Hong Kong could be attributable to ozone and \pm levels, respectively. This represents 8-20\% and 4-9\% of the annual number of global visits, respectively. A meta-analysis of national asthma incidence and prevalence rates led to a new dataset of asthma ERV rates produced from survey data in these countries, with east and south Asian countries accounting for more than half the asthma ERVs. \pm and ozone concentrations from remote sensing satellites were used to estimate the pollution-attributable fraction of asthma ERVs. A more recent study analysed the short-term impact of different \pm levels on the use of rescue inhalers by asthmatics \cite{b_7}. The pollution data were obtained from an air quality monitoring station (AQMS) in the same state. The study found an increase of less than 1\% in medication usage for a corresponding 12\% increase in \pm concentration, however the average distance of the AQMS from the monitored subjects was 16km - the pollutant concentrations recorded were unlikely to be in line with the subjects' true personal exposure. The same problem was identified in \cite{b_8} which demonstrated that sparsely located air quality monitoring stations are not sufficient to identify all the pollutant concentration peaks in an area. By installing community sensors in a region of California the authors were able to identify twice the number of high air pollution episodes compared to using the air quality monitoring stations alone. We address this issue in our work by measuring the \pm concentration where it really matters: on the person, thanks to the wearable Airspeck device.

Yabin et al. \cite{b_9} investigated the association between meteorological conditions and air pollution and incidence of childhood allergic diseases such as asthma, allergic rhinitis and atopic dermatitis based on hospital admissions data of around 2.5 million children in two paediatric hospitals in Shanghai over an 11 year period. Daily averages of meteorological data (temperature, relative humidity, air pressure, precipitation, radiation and wind speed) were obtained from the Shanghai Meteorological Centre. In the absence of residential addresses of the study subjects, air pollution data (PM$_{10}$, PM$_{2.5}$, NO$_2$, SO$_2$ and O$_3$) was averaged over readings obtained from a network of air quality monitoring stations distributed widely across the city, operated by the Shanghai Environmental Protection Agency. The study found that climatic variations in the daily mean temperature and air pressure, more so than changes in air pollution, were associated with the occurrence of childhood allergic diseases. The advances in our work include the use of high resolution, minute level exposure data for \pm (and temperature and relative humidity) obtained from the wearable Airspeck sensor, combined with minute level recordings of respiratory rate and activity level from the Respeck sensor. This addresses the weaknesses of previous work which assumes constant exposure levels over distances of several kilometers which has been demonstrated not to be the case \cite{b_8}. Additionally, previous work has investigated responses which are discrete significant events such as visits to the ER, hospitalisations due to exacerbations or even mortality in the extreme. In practice asthmatics experience a continuum of responses to irritation ranging from discomfort to the use of an inhaler to being laid low. Changes in respiratory rate due to air pollution exposure provide a more subtle means of gauging the health effects of air pollution in vulnerable groups such as those with asthma or COPD.

\section{Causal Discovery}
A causal model of a system entails a probability model along with information regarding the causal relationships between variables. In the same sense that statistical learning is the problem of estimating properties of the underlying distribution given empirical evidence, causal discovery is the problem of estimating properties of the causal model given empirical evidence. In recent years due to access to both large datasets and large computational resources, algorithms have emerged which estimate causal graphs as shown in Figure \ref{fig:causal_graph} from empirical evidence (which can be observational, experimental or both). Such algorithms are called causal discovery algorithms and our approach to establishing exposure-response relationships relies on their use.

\subsection{Assumptions} \label{subsec:assumptions}
Causal discovery, much like statistical learning, is an ill-posed problem \cite{vapnik}: it is impossible for a finite sample of data to encode all of the information of its underlying distribution, let alone its interventional distributions. In both settings we must include \textit{a priori} knowledge in the form of assumptions to allow for identifiability. We now outline the main assumptions of causal discovery required for our approach.

Let $\mathcal{G}$ be a causal graph. A path in $\mathcal{G}$ is any sequence of two or more distinct vertices $x_1, \ldots, x_p$ such that there is an edge between $x_i$ and $x_{i+1}$ for all $i = 1, \ldots, p-1$. If in this path there exist edges $x_i \xrightarrow[]{} x_{i+1}$ and $x_{i+2} \xrightarrow[]{} x_{i+1}$ we say that $x_{i+1}$ is a collider relative to this path. If every path between two nodes $x_i$ and $x_j$ is blocked by a collider we say they are d-separated. The vast majority of causal discovery algorithms work by establishing a link between d-separations in the causal graph and conditional independencies within the data. This link is constructed via the Causal Markov Condition and the Faithfulness assumption \cite{cps}. Let $\mathcal{X}$ be a set of variables with causal relationships between them, $\mathbb{P}_{\mathcal{X}}$ be their joint distribution and $\mathcal{G}$ be their causal graph. The Causal Markov Condition requires that every variable $x \in \mathcal{X}$ be statistically independent of its non-effects given its causal parents $\mathcal{P}(x)$ - the set of direct causes of $x$. Note that this assumption can be considered a generalisation of the well-known Reichenbach's Common Cause Principle \cite{reichenbach}. In practice it allows one to read conditional independence relations directly from the causal graph. However it does not immediately follow that \textit{every} statistical independence can be read from $\mathcal{G}$. We say that $\mathbb{P}_{\mathcal{X}}$ is Faithful to $\mathcal{G}$ if every statistical independence between members of $\mathcal{X}$ is represented by d-separation in $\mathcal{G}$. In conjunction these two assumptions enable a causal discovery algorithm to construct a causal graph which matches the independence relations in the empirical evidence.

The claim that carrying a lighter in one's pocket increases their risk of lung cancer illustrates a common source of misclassified causal relationships - the common cause of both variables is smoking. Such a common cause is known as a confounder and some causal discovery algorithms require all confounders of any two or more variables in $\mathcal{X}$ to either be included in $\mathcal{X}$ or constant for all units in the population. This additional requirement is known as the Causal Sufficiency Assumption.

\subsection{Time Series} \label{subsec:time_series}
Causal discovery has at its core a relationship with time since any effect must follow its cause. It is often the case that our empirical evidence is in the form of a multivariate time series: for the sake of this example let our set $\mathcal{X}$ consist of $d$ variables defined over a time index set $\mathcal{T} \subseteq \mathbb{Z}$, writing $\mathcal{X} := \{X^1_t, X^2_t, \ldots, X^d_t\}_{t \in \mathcal{T}}$. Any causal relationship in $\mathcal{X}$ must be of the form $X^i_{t-\tau} \xrightarrow[]{} X^j_{t}$ with $0 \leq \tau \leq \tau_{max}$ for some maximum possible delay between cause and effect $\tau_{max}$, which could in theory be infinite\footnote{Note that we allow contemporaneous (instantaneous) causal relationships of the form $X^i_{t} \xrightarrow[]{} X^j_{t}$ however in this case we must have $i \neq j$. In practice these represent causal relationships at a delay shorter than the sampling rate of our data.}. Two causal graphs are defined, both of which describe the system:
\begin{itemize}
    \item The full causal graph is the (potentially infinite) DAG $\mathcal{G} := (V, E)$ with vertices $V = \{X^1_t, X^2_t, \ldots, X^d_t\}_{t \in \mathcal{T}}$. We include an edge for every direct causal relationship $X^i_{t-\tau} \xrightarrow[]{} X^j_{t}$ in the system. Note that Figure \ref{fig:causal_graph} is an example of a full causal graph.
    \item The summary causal graph is the finite DAG $\mathcal{G'} = (V', E')$ with vertices $V' := \{X^1, X^2, \ldots, X^d\}$. We include an edge $e = X^i \xrightarrow[]{} X^j \in E'$ if there exists a direct causal relationship $X^i_{t-\tau} \xrightarrow[]{} X^j_{t}$ for any valid $\tau$.
\end{itemize}
Many causal discovery algorithms for time series operate under the Causal Stationarity assumption in addition to the assumptions defined above. We say the time series process $\mathcal{X}$ is causally stationary over the time index set $\mathcal{T}$ if and only if for all causal links $X^{i}_{t-\tau} \xrightarrow{} X^{j}_{t}$ in the system 
\begin{align*}
    X^{i}_{t-\tau} \not{\perp} X^{j}_{t} | \mathbf{X}^{-}_{t} \setminus \{X^{i}_{t-\tau}\} \text{ holds for all } t \in \mathcal{T},
\end{align*}
where $\mathbf{X}^{-}_{t}$ denotes the entire history of the time series process with present observations of other variables. This assumption is weaker than more well-known definitions of stationarity, such as stationarity in mean or variance. Indeed even the strength of the causal relationships themselves may vary over time - the only requirement here is that the existence of the relationships, i.e. conditional independence, remains stationary.

\subsection{Constraint- vs Score-Based Algorithms}
Modern causal discovery algorithms come in two types \cite{score_constraint}: constraint- and score-based algorithms. Constraint-based algorithms test for conditional independencies in the data and construct a causal graph which entails all and only such independencies. Note however that different causal graphs can entail the same probability distribution, so without further assumptions constraint-based methods can only identify the set of possible causal graphs, known formally as the Markov Equivalence Class. One of the main advantages of using these methods is their flexibility: they can be paired with virtually any statistical test for conditional independence. For example in \cite{b_18} the author develops a fully non-parametric test for conditional independence using an estimator of conditional mutual information (a closely related information-theoretic concept) based on the $k$ nearest neighbours algorithm. Such a test is suited to continuous data with complex nonlinear dependencies. Examples of constraint-based algorithms include PC \cite{b_16}, FCI \cite{cps}, PCMCI \cite{b_14} and PCMCI+ \cite{pcmciplus}. Score-based methods instead search over the space of possible causal graphs, seeking to maximise some defined score which reflects how well a given graph fits the data. This score is usually tied to the likelihood of a graph $\mathcal{G}$ given the data $\mathcal{D}$, $\mathbb{P}(\mathcal{D} | \mathcal{G})$. One of the main advantages of these methods is that they can provide a clear measure of confidence in their output. Examples of score-based methods include DYNOTEARS \cite{dynotears} and VARLiNGAM \cite{varlingam}. The latter is an example of an algorithm which enforces further constraints - it assumes a standard vector autoregressive model for the time series process with non-Gaussian noise terms.

\subsection{PCMCI/PCMCI+}
PCMCI and PCMCI+ are recent examples of constraint-based causal discovery algorithms that are particularly suited to multivariate time series, are computationally efficient and have good false-positive control. PCMCI is a two-stage process. The first stage (the PC stage) is a Markov set discovery algorithm used to estimate a superset of the causal parents of every variable in the system through a modified version of the PC algorithm. The pseudocode is given in Algorithm \ref{alg:pcmci} with a brief explanation below. For more detailed pseudocode and a complete description we refer the reader to \cite{b_14}.

\IncMargin{1em}
\begin{algorithm}
\SetKwData{Maxlag}{max\_lag}
    \SetKwInOut{Input}{input}\SetKwInOut{Output}{output}
    \Input{variable $X^{i}_{t}$ to evaluate, maximum lag length \Maxlag, conditional independence test function}
    \Output{superset of causal parents of $X^{i}_{t}$, $\hat{\mathcal{P}}\left(X^{i}_{t}\right)$}
    \BlankLine
    $\hat{\mathcal{P}}\left(X^{i}_{t}\right) = \{\mathbf{X}_{t-1}, \mathbf{X}_{t-2},\ldots,\mathbf{X}_{t-\Maxlag}\}$\;
    \For{$q \leftarrow 0$ \KwTo $|\hat{\mathcal{P}}\left(X^{i}_{t}\right)|$}{
        sort $\hat{\mathcal{P}}\left(X^{i}_{t}\right)$ based on chosen test statistic\;
        $\mathcal{S} =$ first $q$ parents in $\hat{\mathcal{P}}\left(X^{i}_{t}\right)$ based on sorting\;
        \For{$X^{j}_{t-k} \in \hat{\mathcal{P}}\left(X^{i}_{t}\right)$}{
            test $H_0: X^{j}_{t-k} \not\perp X^{i}_{t}|\mathcal{S}$\;
            \If{$H_0$ is not rejected}{
                $\hat{\mathcal{P}}\left(X^{i}_{t}\right) = \hat{\mathcal{P}}\left(X^{i}_{t}\right) $\textbackslash$ \{X^{j}_{t-k}\}$;
            }
        }
    }
    \Return $\hat{\mathcal{P}}\left(X^{i}_{t}\right)$\;
    \caption{PCMCI: First Stage \newline Multivariate time series $\mathbf{X} = \left(X^1, X^2,\ldots,X^N\right)$ of $N$ variables.}\label{alg:pcmci}
\end{algorithm}
\DecMargin{1em}

The intuitions motivating Algorithm \ref{alg:pcmci} are as follows:
\begin{itemize}
    \item The algorithm begins by initialising a superset of all possible causal parents of $X^{i}_{t}$.
    \item The outer loop repeats for all possible subset sizes of $\hat{\mathcal{P}}\left(X^{i}_{t}\right)$. Note that $|\hat{\mathcal{P}}\left(X^{i}_{t}\right)| = N \cdot $ max\_lag.
    \item PCMCI is fast because it sorts $\hat{\mathcal{P}}\left(X^{i}_{t}\right)$ based on the chosen statistic and only tests the $q$ parents with strongest dependency instead of every possible subset $\mathcal{S}$ with $|\mathcal{S}| = q$. The test itself is for conditional independence - if the null hypothesis cannot be rejected at an appropriate significance level the link can be removed from $\hat{\mathcal{P}}\left(X^{i}_{t}\right)$.
    \item In this sense the algorithm recursively removes links it is confident enough to remove in order to obtain the superset $\hat{\mathcal{P}}\left(X^{i}_{t}\right)$ of the causal parents of $X^{i}_{t}$.
\end{itemize}

Algorithm \ref{alg:pcmci} is repeated for every variable $X^{i}_{t} \in \mathbf{X}_t$.

In the second stage each remaining link is tested for momentary conditional independence (MCI). This is an extension of the standard test for conditional independence which conditions on the causal parents of \textit{both} variables in the link. That is, for a link $X^i_{t-p} \rightarrow X^j_t$ the MCI test is
\begin{align*}
    X^i_{t-p} \perp X^j_t | \hat{\mathcal{P}}\left(X^j_t\right) \setminus \{X^i_{t-p}\}, \hat{\mathcal{P}}\left(X^i_{t-p}\right).
\end{align*}
The additional conditioning controls for highly autocorrelated data which leads to much lower false-positive rates than other commonly-used causal discovery methods. The significance of each link can finally be assessed from the p-values of this test, with rejection of the null favouring the existence of a causal relationship. 

For linear dependencies the conditional independence test $X \perp Y | Z$, is carried out using partial correlation. This is achieved by fitting two multivariate regression models to predict both $X$ and $Y$ given $Z$. The correlation of the residuals is then evaluated, with the test assuming these residuals are approximately normally distributed.

The test used for nonlinear dependencies is a fully nonparametric test based on conditional mutual information (conditional independence implies this is zero). The test is fully described in \cite{b_18} and briefly explained here. The following estimator for conditional mutual information is used:
\begin{align*}
    \widehat{I}(X ; Y | Z)=\psi(k)+\frac{1}{n} \sum_{i=1}^{n}\left[\psi\left(k_{i}^{z}\right)-\psi\left(k_{i}^{x z}\right)-\psi\left(k_{i}^{y z}\right)\right]
\end{align*}
where $\psi$ is the Digamma function $\psi(x)=\frac{d}{d x} \ln \Gamma(x)$, $n$ is the sample length and $k$ is a parameter specifying the number of nearest neighbours of each sample to be taken from the joint sample space. The main advantage of this nearest-neighbour estimator approach is that it can capture almost any nonlinear dependency - in this sense the test is fully nonparametric. Note however that three separate nearest-neighbour calculations have to be carried out for each sample. This increases runtime considerably compared to the linear test.

PCMCI+ improves on the performance of PCMCI and enables the detection of contemporaneous as well as lagged causal dependencies. For an understanding of this extension we refer the interested reader to \cite{pcmciplus}. We apply PCMCI+ in Section \ref{sec:results} and present results for both a linear and nonlinear conditional independence test for data at both 1 and 15 minute resolutions, up to a maximum lag of 8 hours. PCMCI+ was implemented using the \textit{Tigramite} \cite{b_17} package in Python (3.8).

\section{Results} \label{sec:results}
PCMCI+ was applied to a dataset of 113 asthmatic subjects who were each monitored for 48-hour periods using both the Respeck and Airspeck devices. Each subject's data was recorded as a multivariate time series consisting of minute-level observations on \pm exposure, temperature, relative humidity, breathing rate and intensity of activity\footnote{Intensity of activity was measured using the Respeck device.}. We believe such a complex system can be described by multiple overlapping subsystems, some of which may be characterised by linear causal dependencies and some by nonlinear causal dependencies. We therefore apply the algorithm separately using both a linear and nonlinear test for conditional independence, and compare both sets of results. Note that the former test adds an additional parametric assumption to those discussed in Section \ref{subsec:assumptions}, while the latter adds no new assumptions and is fully nonparametric - at the cost of computational complexity (see Section \ref{sec:discussion}). Due to this cost we resample each time series to a 15-minute resolution when using this nonlinear test. To control false positives due to multiple hypothesis testing, we set a threshold for statistical significance at $p < 0.02$ for the linear test and $p < 0.05$ for the nonlinear test (as the resampled data results in fewer tests overall). 

We select a maximum lag-length $\tau_{max}$ after discussions with clinical experts and viewing unconditional dependencies, focusing on cross-correlations in particular as PCMCI+ is known to perform well even with highly autocorrelated time series \cite{pcmciplus}. After initial experiments we find that unconditional dependencies between variables all decay roughly after 6-8 hours - we therefore set $\tau_{max}$ equal to 8 hours in all experiments. 

\begin{figure}[ht]
    \centering
    \includegraphics[width=0.6\textwidth]{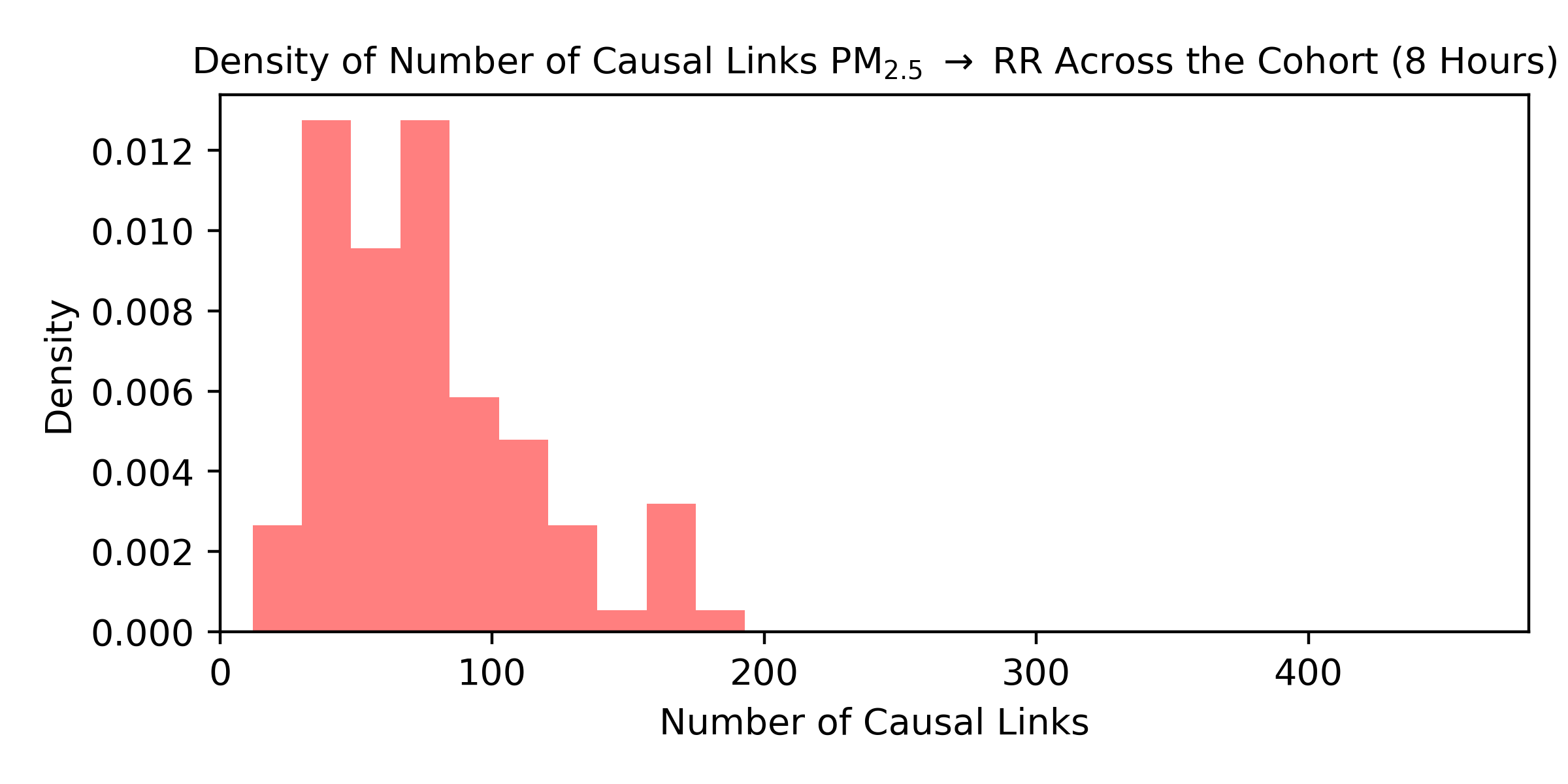}
    \caption{We count the number of direct causal links from lags of \pm to the current respiratory rate for all subjects. Since our data is sampled every minute up to a maximum delay of 8 hours, this is a non-negative integer less than or equal to 481 (including contemporaneous effects). We note significant variation between subjects, however most subjects have a relatively low number of causal links.}
    \label{fig:ncl_density}
\end{figure}

After running PCMCI+ on a given subject's time series we can count the number of direct causal links of the form \pm($t-\tau$) $\rightarrow$ (Breathing Rate)($t$) for each lag $\tau$ up to $\tau_{max}$, and compare these totals between other subjects. Figure \ref{fig:ncl_density} shows the distribution of the number of causal links of this form over the entire cohort over the entire 8 hours, for our experiments using a linear test for PCMCI+ at a one-minute resolution. Note that the distribution is not narrow i.e. there is much variation in the number of causal links between subjects, suggesting the study of exposure-response relationships for this effect should be done at the personal level. We do see however that the majority of subjects have a relatively low number of causal links when compared to the maximum lag. This indicates that while direct causal effects from \pm exposure to breathing rate exist, they are not drawn out over a period of several hours at a time. Therefore it would be productive to examine whether such causal effects are more likely to occur after certain delays than others.

\begin{figure}[ht]
    \centering
    \includegraphics[width=0.6\textwidth]{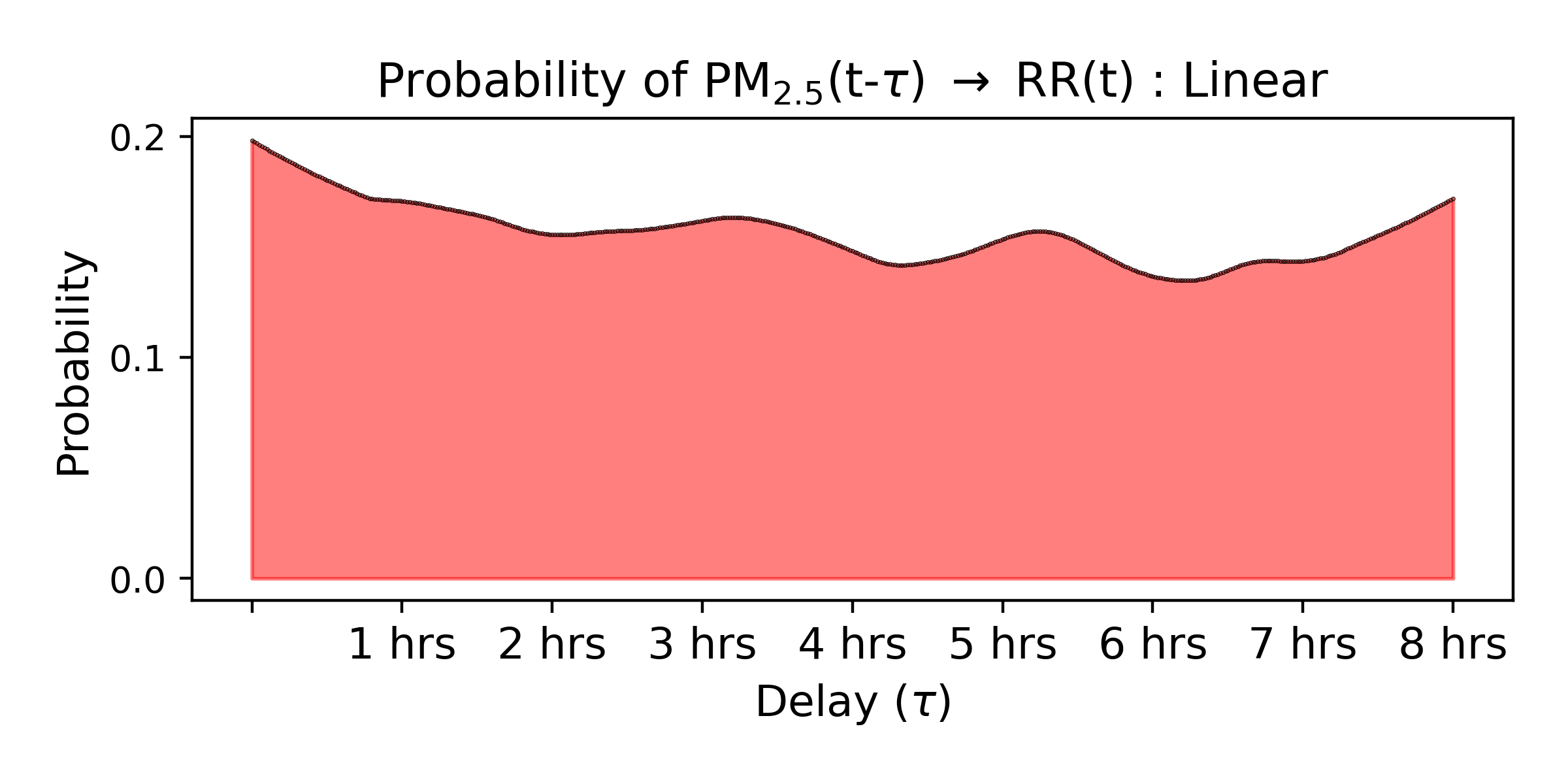}
    \caption{The probability of seeing a causal relationship from a given lag of \pm to the current breathing rate across the cohort, up to the maximum lag of 8 hours. This stays low for all lags, and changes little in this experimental setup, with a slightly higher probability at both the start and end of the 8 hour period.}
    \label{fig:r1t480_pm_8hrs}
\end{figure}

For the same set of experiments, we calculate the proportion of the cohort with a causal relationship \pm($t-\tau$) $\rightarrow$ (Breathing Rate)($t$) for each $\tau$. In this sense we obtain a probability of seeing a causal link at a given lag, and these probabilities are graphed in Figure \ref{fig:r1t480_pm_8hrs}.

Figure \ref{fig:r1t480_pm_8hrs} shows a low ($< 0.2$) probability of seeing a causal relationship from any lag of \pm to the current breathing rate, which matches the low totals shown in Figure \ref{fig:ncl_density}. This probability stays relatively consistent, with perhaps a slightly higher chance of seeing a causal link within the first hour or during the last hour of the 8 hour period. Note that the conditional independence test used here assumes linear dependencies - it may therefore have low power in detecting some nonlinear causal mechanisms in this system. 

\begin{figure}[ht]
    \centering
    \begin{subfigure}{0.6\textwidth}%
        \includegraphics[width=\textwidth]{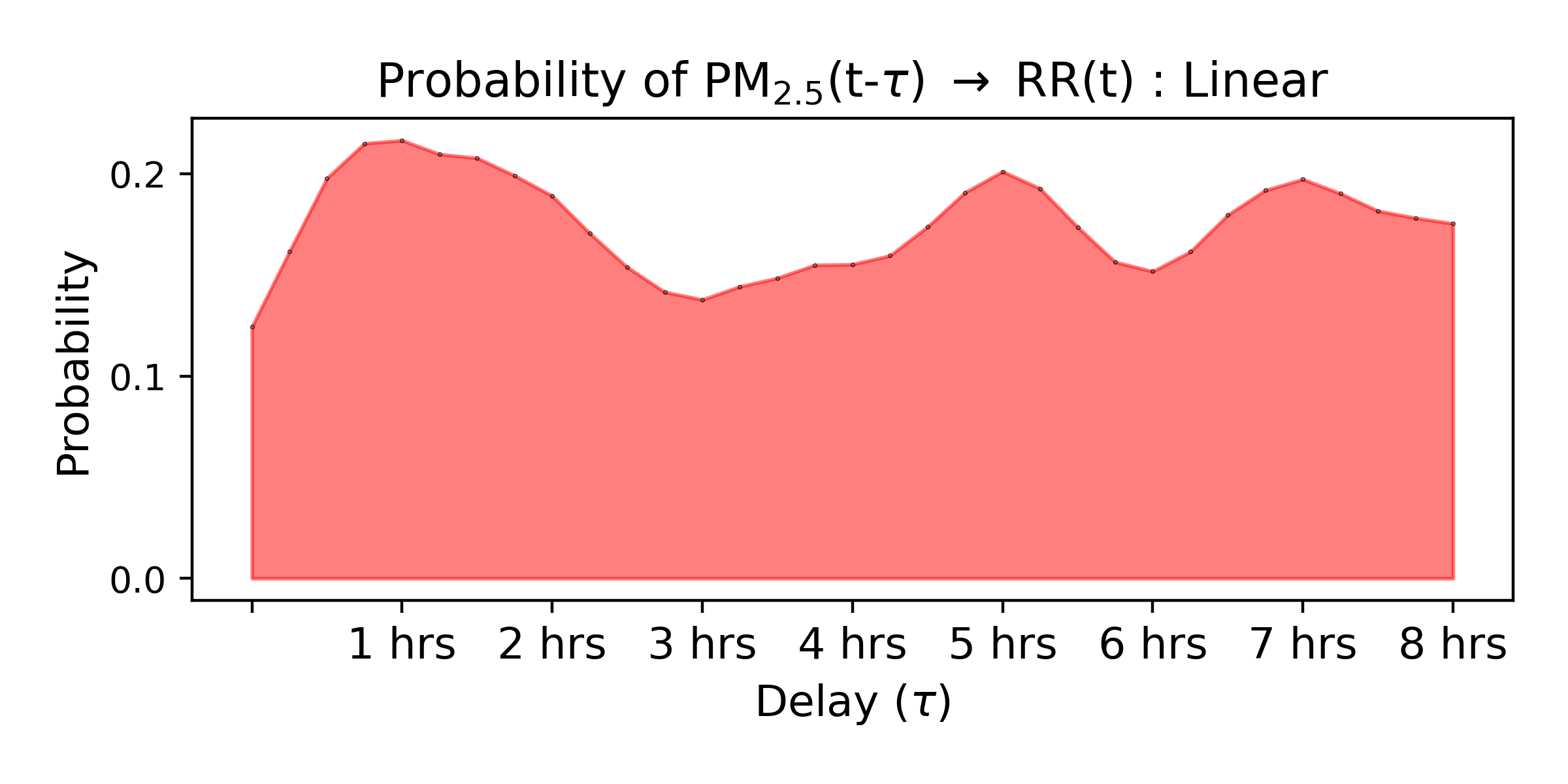}%
    \end{subfigure} \\
    \begin{subfigure}{0.6\textwidth}%
        \includegraphics[width=\textwidth]{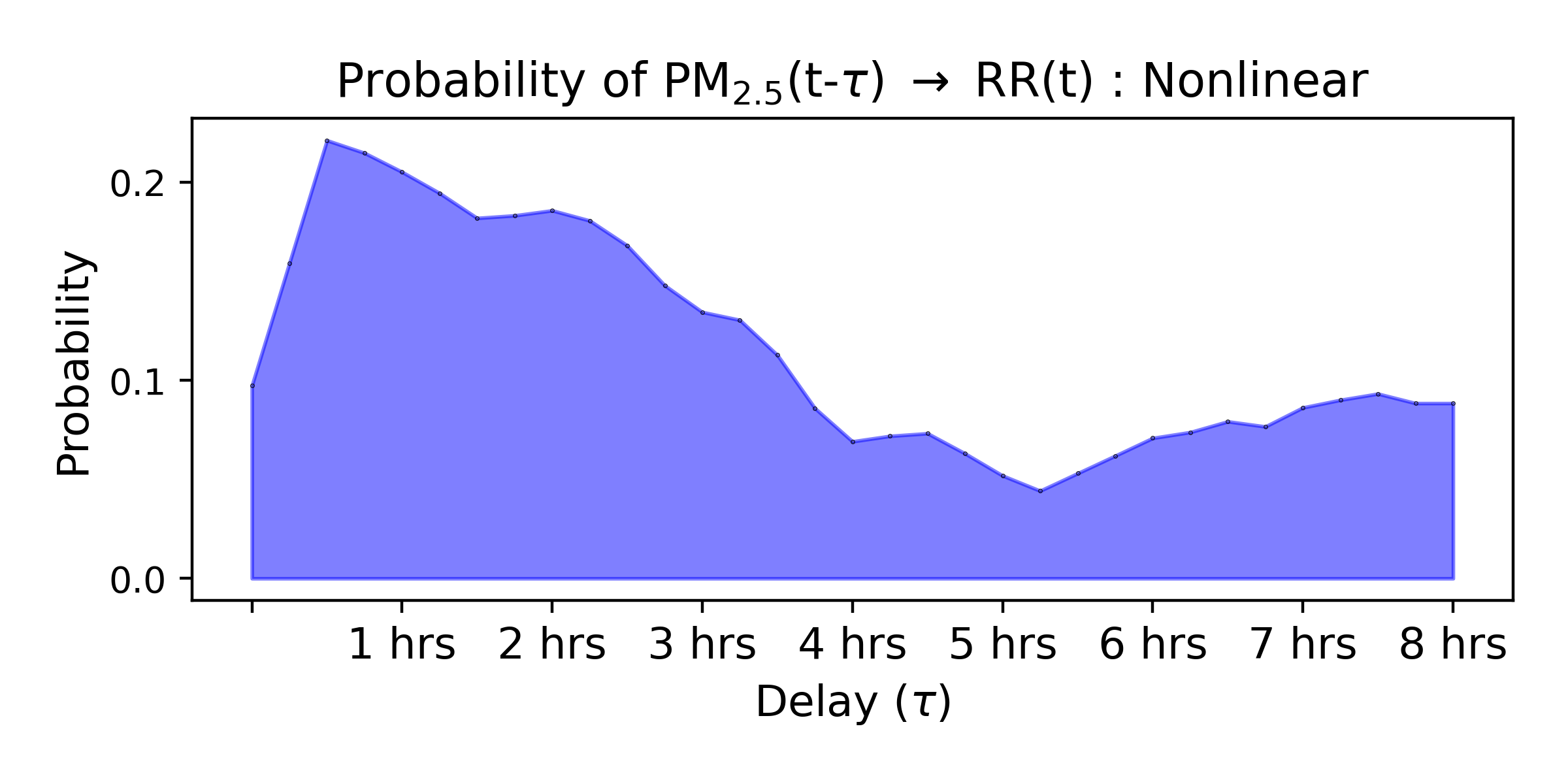}%
    \end{subfigure}%
    \caption{The probability of seeing a causal relationship from a given lag of \pm to the current breathing rate across the cohort, up to the maximum lag of 8 hours. \textbf{Upper}: Linear dependencies. \textbf{Lower}: Nonlinear dependencies. We see a much lower chance of nonlinear dependencies between delays of roughly 3-5 hours than linear dependencies.}
    \label{fig:r15t32_pm_split}
\end{figure}

We study such nonlinear mechanisms in an adjusted experimental setup in which we resample our time series to a 15 minute resolution. While this may results in some loss of information, the reduction required runtime for causal discovery allows us to use both a linear and nonlinear conditional independence test and compare the results. Figure \ref{fig:r15t32_pm_split} shows the probability of seeing both linear and nonlinear causal dependencies between lags of \pm and the current breathing rate in this set of experiments. Note that there may be some overlap between these two tests e.g. some nonlinear dependencies may be easy to detect via a first-order approximation when using a linear test. Regardless, they give us some indication of different subsystems in our data as the two graphs differ greatly.

Both graphs also show an overall low probability of seeing causal links. The upper graph for linear dependencies has three clear peaks of roughly similar height: one for early lags around 1 hour, and two more for later lags around 5 hours and 7 hours. At other lags, we see a moderately lower chance of a causal link, comparable to the drop seen in Figure \ref{fig:r1t480_pm_8hrs}. Meanwhile the lower graph for nonlinear dependencies sees a sharp peak within the first hour of lags, followed by a steady and large decrease. After a delay of around 5 hours, the probability of seeing a nonlinear causal dependency steadily begins to rise again up to the maximum delay of 8 hours. These results may hint at an immediate reaction to \pm exposure reflected in the respiratory rate, followed by a longer-term reaction some hours later. It is clear from these two graphs that for real data from complex systems, it is worth studying different types of dependencies as subsystems. 

\begin{figure}[ht]
    \centering
    \includegraphics[width=0.7\textwidth]{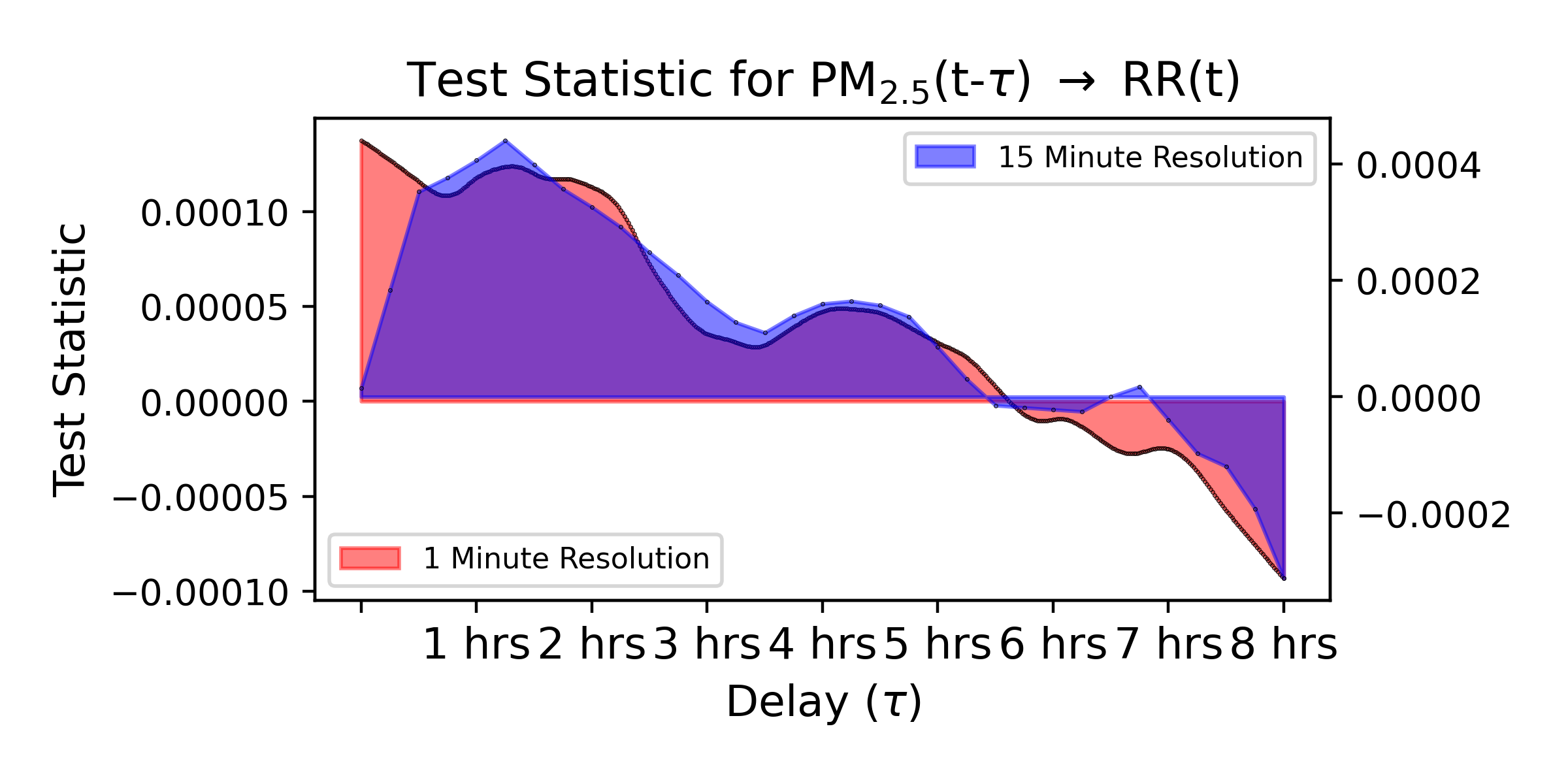}
    \caption{PCMCI+ was run using a linear conditional independence test for a maximum lag of 8 hours at both a 1 and 15 minute resolution. When comparing the average (normalised) test statistics for every lag we see both experimental setups show very similar trajectories, suggesting that much of the information is conserved between them despite resampling.}
    \label{fig:val_comparison}
\end{figure}

Discrepancies in the trajectories of Figures \ref{fig:r1t480_pm_8hrs} and \ref{fig:r15t32_pm_split}-U may be due to a few reasons. It may be the case that resampling to a 15 minute resolution results in a loss of information and therefore loss of power in the statistical tests used (resulting in false negatives). On the other hand it may also be the case that the large number of hypothesis tests carried out when using data at a minute-level resolution result in too many false positives, inflating the number of causal links. We believe the second possibility is more likely, as evidenced by figure \ref{fig:val_comparison}. Instead of examining the presence of causal links, inferred after calculating a p-value, we instead plot the test statistics themselves. In the case of both 1 and 15 minute resolution time series, we run PCMCI+ using a test for linear dependencies and plot the average (normalised) test statistic for each lag in both experimental setups, for the causal link \pm($t-\tau$) $\rightarrow$ (Breathing Rate)($t$) (for each $\tau$). While on different scales, we see that the relative trajectories for both setups are highly correlated. Indeed the main discrepancies between the two trajectories occur for lags less than an hour, and for lags between 6 and 7 hours; this is also clear on examination of Figures \ref{fig:r1t480_pm_8hrs} and \ref{fig:r15t32_pm_split}-U. This is evidence that the difference in resolution between these two setups does not result in the destruction or loss of too much information, and that the calculated p-values which follow are themselves leading to discrepancies. It is likely that false positive rates are inflated in our first set of experiments due to the large number of hypothesis tests carried out.

\section{Discussion} \label{sec:discussion}
The main purpose of any epidemiological study is the examination of distributions and determinants of health-related events in a target population, followed by recommendations for tackling the corresponding health issues. Analysis through machine learning models tends to focus solely on the distributions at the expense of the determinants. Their primary concern is accurate predictions on unseen data rather than an understanding of the data themselves and the causal relationships between features \cite{ml_in_epi}. While these techniques are becoming increasingly popular in published literature, machine learning models – especially deep learning models – are difficult to interpret. For example, high dimensional, high resolution time series data often require heavy preprocessing and the use of dimensionality reduction techniques such as Principle Component Analysis (PCA) to engineer features with high predictive power. These are then used with complex architectures like recurrent neural networks to produce highly accurate forecasts, however the impact of the individual original features is often completely obscured at this point.
In contrast, causal discovery methods generally require no feature engineering as they accept a multivariate time series directly in order to learn the causal relationships between all measured variables.
These methods promise a better understanding of the often subtle and complex dependencies in data gathered in epidemiological studies. This causal structure can also be used to infer the results of distributional shifts or interventions: this knowledge is important when developing methods of control in epidemiology, a task difficult to complete via standard correlation or regression-based approaches due to ambiguity and potential confounders \cite{b_14}.

This paper has examined the causal relationships from personal \pm exposure to changes in respiratory rate. The use of this continuous response is an improvement on previous research which investigated responses in the form of significant discrete events such as admissions to emergency rooms, mortality \cite{b_19}, or relief inhaler usage \cite{b_7}. The \pm exposure and respiratory rate of a given participant was sampled every minute by the wearable Airspeck and Respeck devices respectively, over a period of 48 hours. The resulting time series was processed using a causal discovery algorithm to learn its causal structure.

Several constraint- and score-based causal discovery algorithms exist for use with multivariate time series \cite{score_constraint}. The system under study in this paper is complex: the interactions between personal \pm exposure, temperature, humidity, and the subject’s respiratory rate in unconstrained environments have not been studied before and are not well understood. The chosen causal discovery algorithm should be capable of identifying relationships regardless of whether they are linear or nonlinear mechanically. Constraint-based algorithms such as PCMCI and PCMCI+ are specifically designed to be paired with any test for conditional independence, such as partial correlation for linear causal links and conditional mutual information for nonlinear links. The latter test is fully nonparametric, enforcing no additional assumptions on the data beyond those required to apply the PCMCI and PCMCI+ algorithms \cite{kcmi}.

The \textit{Tigramite} Python package, required to run both PCMCI and PCMCI+, includes implementations for both of these tests. Both do not enforce any strict assumptions on the data themselves beyond those discussed in Sections \ref{subsec:assumptions} and \ref{subsec:time_series}. This presents a clear advantage over other popular causal discovery algorithms such as VARLiNGAM which at all times assumes a standard vector autoregressive model. PCMCI+ was preferred as it tests for contemporaneous causal relationships between variables (unlike PCMCI) – these can be interpreted as causal links with a delay shorter than the sampling rate of the time series.

Running PCMCI+, and indeed many causal discovery algorithms, requires the Causal Sufficiency assumption. This can be a challenge to enforce and often requires expert domain knowledge to consider possible confounders of the relationship under study. In this paper the measurement of potential confounders and the choice of an appropriate maximum lag-length, $\tau_{max}$, were based on discussions with clinical experts. This led to the inclusion of confounders such as temperature and the decision to set a maximum lag-length of 8 hours for delayed effects of exposure.

The main complexity of PCMCI+ is in the multiple conditional independence tests carried out: in particular, the nonlinear test of conditional mutual information requires the construction of a k-d tree and three search operations for every single test (average search complexity is O(log n) and worst-case search complexity is O(n)) \cite{kcmi}. The execution time of the algorithm on large time series can be on the order of days when run on a high-end laptop such as a MacBook Pro 2020. A high-performance Linux computing cluster (consisting of over 7000 Intel Xeon cores with up to 3TB memory available on a single node) was employed to parallelize individual experiments and reduce the execution time of a single experiment to around 6 hours. 

\begin{figure}[ht]
    \centering
    \includegraphics[width=0.56\textwidth]{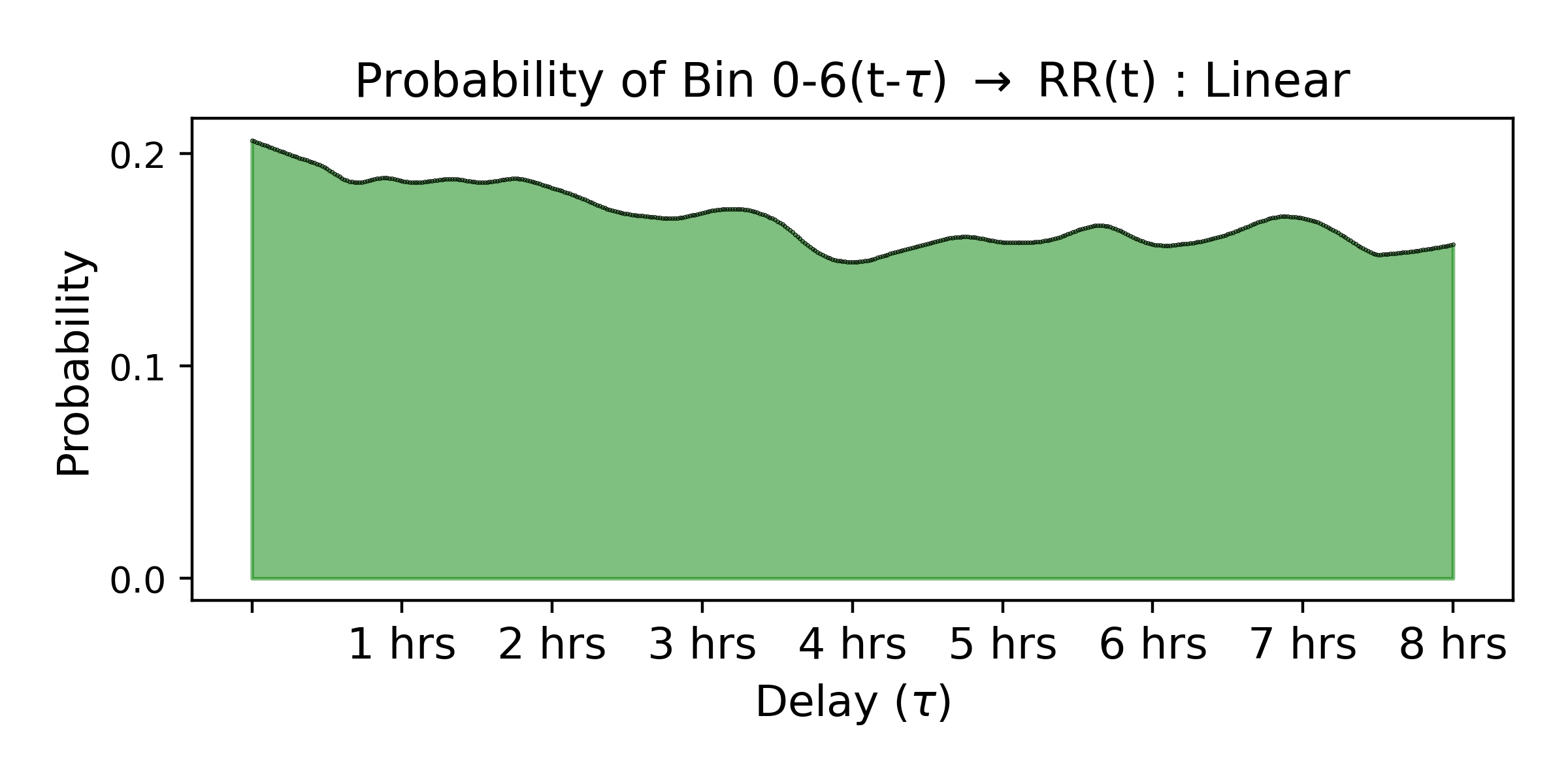}
    \caption{The probability of seeing a causal relationship from a given lag of bins 0-6 to the current breathing rate across the cohort, up to the maximum lag of 8 hours. The magnitude of these probabilities, as well as the general trajectory over all lags, is similar to those of Figure \ref{fig:r1t480_pm_8hrs}.}
    \label{fig:r1t480_bins_8hrs}
\end{figure}

Sensor data are often noisy and require calibration before comparing effect sizes. This is another complexity in similar data-driven studies. The Airspeck sensor works by counting particles of different diameters and updating a running total of 16 distinct (arbitrary) bins, numbered 0-15. For \pm, bins 0-6 all count particles of diameter $\leq 3\bm{\upmu}$m and therefore constitute a superset of \pm. A nonlinear transformation is required to calculate a concentration from the particle counts of bins 0-6, which then must be calibrated to a reference monitor for consistency between different sensors. This time-consuming task was carried out in order to work with \pm measurements as these are widely recognised standard measures of air quality. For causal identification however, we may not require calibration between sensors as we are not concerned with effect sizes. We investigate this empirically by applying causal discovery directly to bins 0-6 instead of \pm measurements, with the hope of obtaining comparable results as the bins are in a sense ``root causes" of \pm. As in Section \ref{sec:results}, we evaluate the probability of finding a causal link from a given lag of bins 0-6 to the current breathing rate, and plot these probabilities in Figure \ref{fig:r1t480_bins_8hrs}. While the magnitude of these probabilities and the general trajectory across all lags is comparable to Figure \ref{fig:r1t480_pm_8hrs}, it is difficult to assess from this graph whether the use of relevant bin counts is an appropriate substitution for \pm in terms of shared information, as Figure \ref{fig:r1t480_bins_8hrs} shows cohort-wide proportions. We know from the results in Section \ref{sec:results} that individual causal effects may differ greatly between subjects, so we instead assess the correlation between the calculated p-values for causal links of the form \pm($t-\tau$) $\rightarrow$ (Breathing Rate)($t$) and those of the form (Bin $i$)($t-\tau$) $\rightarrow$ (Breathing Rate)($t$) for $i \in \{0 - 6\}$. We do this for all subjects across the cohort.

\begin{table}
\centering
\resizebox{.3\textwidth}{!}{%
\begin{tabular}{l|ll}
    \toprule
    & \textbf{$r$} & \textbf{$\rho$} \\
    \midrule 
    \textbf{Bin 0} & 0.114 & 0.138\\
    \textbf{Bin 1} & 0.17 & 0.208\\
    \textbf{Bin 2} & 0.222 & 0.262\\
    \textbf{Bin 3} & 0.245 & 0.28\\
    \textbf{Bin 4} & 0.194 & 0.217\\
    \textbf{Bin 5} & 0.145 & 0.162\\
    \textbf{Bin 6} & 0.122 & 0.145\\
    \midrule
    \textbf{All Bins} & 0.32 & 0.33 \\
    \bottomrule
\end{tabular}%
}
\vspace{1em}
\caption{Pearson correlation ($r$) and Spearman correlation ($\rho$) between the p-values for causal relationships between \pm and breathing rate compared to those for causal relationships between relevant bin counts and breathing rate, as well as bins 0-6 summed. We see that the magnitude of these correlations is quite low.}
\label{table:pvals}
\end{table}

The results are shown in Table \ref{table:pvals}. We assess both a Pearson (linear) correlation and Spearman (ranked) correlation between the p-values for causal relationships between \pm and breathing rate and between bin counts and breathing rate. We also sum bins 0-6, calculate p-values for causal links and compare the results with \pm. All correlations are low in magnitude: only when all bins are summed do we find a low-moderate correlation of roughly 0.3. These results suggest that bin counts are not a sufficient replacement for identifying causal relationships between lags of \pm exposure and the breathing rate. However, following the methodology presented in Section \ref{sec:results} we assess the correlation between the test statistics themselves instead of the calculated p-values, and present the results in Table \ref{table:vals}. We find that the magnitudes of these correlations are much higher, note particularly that the test statistics for relationships between all bins summed and the breathing rate are highly correlated with test statistics for relationships between \pm and the breathing rate. For reference, we include Table \ref{table:data} which shows the Pearson and Spearman correlation between the time series themselves before any causal discovery has been applied, finding that the magnitude correlations are comparable to Table \ref{table:vals}.

\begin{table}
\centering
\resizebox{.3\textwidth}{!}{%
\begin{tabular}{l|ll}
    \toprule
    & \textbf{$r$} & \textbf{$\rho$} \\
    \midrule 
    \textbf{Bin 0} & 0.545 & 0.543\\
    \textbf{Bin 1} & 0.65 & 0.643\\
    \textbf{Bin 2} & 0.699 & 0.699\\
    \textbf{Bin 3} & 0.71 & 0.715\\
    \textbf{Bin 4} & 0.626 & 0.636\\
    \textbf{Bin 5} & 0.525 & 0.542\\
    \textbf{Bin 6} & 0.435 & 0.451\\
    \midrule
    \textbf{All Bins} & 0.78 & 0.784 \\
    \bottomrule
\end{tabular}%
}
\vspace{1em}
\caption{Pearson correlation ($r$) and Spearman correlation ($\rho$) between test statistics for causal relationships between \pm and breathing rate compared to those for causal relationships between relevant bin counts and breathing rate, as well as bins 0-6 summed. Many correlations are high in magnitude and are indeed comparable to correlations between the data streams themselves (see Table \ref{table:data}).}
\label{table:vals}
\end{table}

\begin{table}
\centering
\resizebox{.3\textwidth}{!}{%
\begin{tabular}{l|ll}
    \toprule
    & \textbf{$r$} & \textbf{$\rho$} \\
    \midrule 
    \textbf{Bin 0} & 0.574 & 0.76 \\
    \textbf{Bin 1} & 0.707 & 0.841 \\
    \textbf{Bin 2} & 0.752 & 0.849 \\
    \textbf{Bin 3} & 0.687 & 0.825 \\
    \textbf{Bin 4} & 0.617 & 0.742 \\
    \textbf{Bin 5} & 0.463 & 0.672 \\
    \textbf{Bin 6} & 0.347 & 0.571 \\
    \midrule
    \textbf{All Bins} & 0.688 & 0.887 \\
    \bottomrule
\end{tabular}%
}
\vspace{1em}
\caption{Pearson correlation ($r$) and Spearman correlation ($\rho$) between the original time series for \pm exposure and relevant bin counts across all subjects in the cohort. We find that the magnitude of these correlations is high and comparable to those seen in Table \ref{table:vals}.}
\label{table:data}
\end{table}

These results further emphasise the concluding remarks of Section \ref{sec:results}: namely that false positive control in causal discovery algorithms like PCMCI+ is a clear bottleneck of this approach to similar epidemiology problems. The large correlations seen in Table \ref{table:vals} suggest that bin counts may be used instead of \pm measurements when researchers care only about causal identification, disregarding effect sizes. This is due to a lot of shared information between the two data streams. The use of 7 variables instead of 1 however likely exacerbates the multiple testing problems seen previously and increases the number of false positive causal dependencies found, leading to unreliable p-values. We therefore stress the importance of future work on false positive control for causal discovery algorithms.

Future research will investigate the pairing of causal discovery methods and predictive modelling techniques for deriving the average causal effect as a function of causal parents. The personal exposure-response relationships discussed in Section \ref{sec:results} could be derived using the learned causal structure to infer a predictive model. One possible way of doing this is to assume an additive structure for a predictive model of breathing rate given measurements of \pm, temperature, humidity and activity level as measured in this paper. Generalised additive mixed models (GAMMs) could then be used to quantify the individual effects of each variable, in particular the personal \pm exposure. The form of the model itself could be inferred using the causal parents of a given subject’s respiratory rate.

The clinical interpretation of the results in Section \ref{sec:results} is currently underway to investigate the apparent greater likelihood of causal relationships within the first hour and after five hours as seen in Figure \ref{fig:r15t32_pm_split}, and the differences in the response of different subjects as highlighted in Figure \ref{fig:ncl_density}.

\section{Conclusion}
This paper has described and demonstrated a method for establishing causal relationships from personal \pm exposure to the respiratory rate for a cohort of asthmatic subjects through the use of a pair of wearable sensors: the Airspeck for measuring \pm exposure and the Respeck for continuously recording the respiratory rate, while controlling for temperature, relative humidity and activity level. For the first time a personalised exposure-response relationship for the short-term effects of air pollution has been established for asthmatics during 48 hours of their everyday lives. Medical interpretation of the distribution of causal relationships found in Figure \ref{fig:ncl_density} is being undertaken with our clinical collaborators.

This approach teases out the impact of \pm exposure at a temporal resolution hitherto impossible: 1 and 15 minute-level resolutions up to the maximum delay of 8 hours. This has been possible thanks to wearable sensors generating time series datasets at such high resolution coupled with high-performance computing-based sensor data analytics. In contrast to more traditional epidemiological approaches the strength of our data-driven approach comes from the large volume of observations which are crunched to derive causal relationships. Another important consequence of this approach is the ability to learn the exposure-response relationship at the individual level rather than at the cohort level. This information can then be used to devise mitigation strategies which are customised for the individual, e.g. planning of clean commuting routes for those vulnerable to deteriorating air quality. 

\section*{Acknowledgments}
This work was supported in part by the UK NERC-/MRC- funded DAPHNE project (NE/P016340); the UK MRC-/AHRC- funded PHILAP project (MC\_PC\_MR/R024405/1) and the UK EPSRC-funded INHALE project (EP/T003189/1).

\section*{Ethics Approval}
The 113 asthmatic subjects were recruited at the All India Institute of Medical Sciences (AIIMS) as part of the DAPHNE study: established to investigate the impact of air pollution on respiratory health in asthmatic adolescents, starting in August 2018 and disrupted by the COVID-19 pandemic in March 2020.  Ethics approval for the DAPHNE study was granted by the Institute Ethics Committee of AIIMS (Reference numbers: IEC-256/05.05.2017, RP-26/2017, OP13/03.08.2018).


\end{document}